\documentclass[12pt,twocolumn]{iopart}
\usepackage{amsmath}
\usepackage{graphicx}
\usepackage{psfig}
\usepackage{cite}

\begin{document}

\title{Inversion formulas for the broken-ray Radon transform}

\author{Lucia Florescu}
\address{Department of Bioengineering, University of Pennsylvania,
  Philadelphia, Pennsylvania 19104}

\author{Vadim A. Markel}
\address{Departments of Radiology and Bioengineering, University
of Pennsylvania, Philadelphia, Pennsylvania 19104}

\author{John C. Schotland} 
\address{Department of Bioengineering and
  Graduate Group in Applied Mathematics and Computational Science,
  University of Pennsylvania, Philadelphia, Pennsylvania 19104}

\begin{abstract}
  We consider the inverse problem of the broken ray transform
  (sometimes also referred to as the V-line transform).  Explicit
  image reconstruction formulas are derived and tested numerically.
  The obtained formulas are generalizations of the filtered
  backprojection formula of the conventional Radon transform. The
  advantages of the broken ray transform include the possibility to
  reconstruct the absorption and the scattering coefficients of the
  medium simultaneously and the possibility to utilize scattered
  radiation which, in the case of the conventional X-ray tomography,
  is typically discarded.
\end{abstract}

\submitto{\IP}
\date{\today} 

\maketitle
\section{Introduction}

Image reconstruction techniques based on inversion of the Radon
transform are well-established. While the classical Radon transform
utilizes straight rays, there exists a considerable interest in the
physical situations wherein the straight-ray propagation is lost due
to the effects of scattering or refraction, yet measurements can be
mathematically related to integrals of the medium properties over
well-defined trajectories. This leads to various generalizations of
the conventional Radon transform. For example, it was shown that the
Radon transform on co-planar circles whose centers are restricted to a
bounded domain is invertible~\cite{agranovsky_96_1}. Radon transforms
on other smooth curves have also been
considered~\cite{maass_89_1,lissiano_97_1,maeland_98_1}.  Recently, a
series of papers have explored a circular-arc transform which arises
when the signal is generated by first-order Compton scattering of
X-rays~\cite{driol_08_1,nguyen_10_1}. In the imaging modality proposed
in references~\cite{driol_08_1,nguyen_10_1}, the contrast mechanism is
related to the spatially-varying efficiency of Compton scattering
while attenuation of the scattered rays by the medium is neglected.
Therefore, the absorption coefficient can not be recovered using this
modality.

We have recently proposed an approach which is also based on the
single-scattering approximation but allows one to take into account
and to reconstruct both the attenuation and the scattering
coefficients of the medium~\cite{florescu_09_1,florescu_10_1}. The
technique is applicable to either Compton scattering in the case of
X-ray tomography, or to elastic scattering in the case of optical
tomography, and was termed by us as SSOT (single-scattering optical
tomography). The mathematical underpinning of SSOT is the broken ray
transform of the medium which is obtained by employing collimated
sources and detectors whose illumination/detection direction vectors
lie in the same plane but are not on axis; the former defines the
slice in which the image is reconstructed. A broken ray consists of
two straight segments connected by a vertex and, generally, resembles
the letter ``V'' (the angle in ``V'' can be larger than $\pi/2$). A
similar transform was considered recently in reference~\cite{morvidone_10_1},
where the physical mechanism of image formation was related to Compton
scattering of gamma-rays emitted by an intrinsic radioactive contrast
agent.

It can be seen that generalizations of the Radon transform to
non-smooth curves are relatively new and unexplored. This is
especially true for the case of SSOT which employs broken rays whose
directions take only few (possibly, only two) discrete values.  This
is in sharp contrast to the conventional idea that multiple ray
directions are required for stable image reconstruction. The fact that
the broken-ray transform is invertible is, therefore,
counterintuitive. Yet, in
references~\cite{florescu_09_1,florescu_10_1}, image reconstruction
was demonstrated using a purely numerical algorithm which involved
discretization of the integral transform and seeking the
pseudo-inverse solution to the resulting system of algebraic
equations. From the singular-value analysis of the discretized
integral transform operator, it was found that the broken ray
transform is mildly ill-posed even in the case when only two ray
directions are used. Moreover, we have shown that, if more than one
broken ray is used for detection, the scattering and the absorption
coefficients of the medium can be reconstructed simultaneously.
Another feature which is potentially useful is that inversion of the
broken ray transform does not require multiple projections and can be,
in principle, performed in the backscattering geometry if
transillumination data are not available.

In this paper, we derive and test numerically explicit image
reconstruction formulas for the broken ray transform. In
section~\ref{sec:BRT}, the transform and some relevant geometrical
quantities are introduced and defined. In section~\ref{sec:img_rec},
the image reconstruction formulas are derived for the cases when the
inhomogeneities are purely absorptive (section~\ref{subsec:abs}) and
when both absorbing and scattering inhomogeneities are present
simultaneously (section~\ref{subsec:abs+scat}). Numerical examples are
given in section~\ref{sec:num}. Finally, section~\ref{sec:sum}
contains a discussion and a brief summary of obtained results.

\section{The broken ray transform}
\label{sec:BRT}

\begin{figure}
\centerline{
\psfig{file=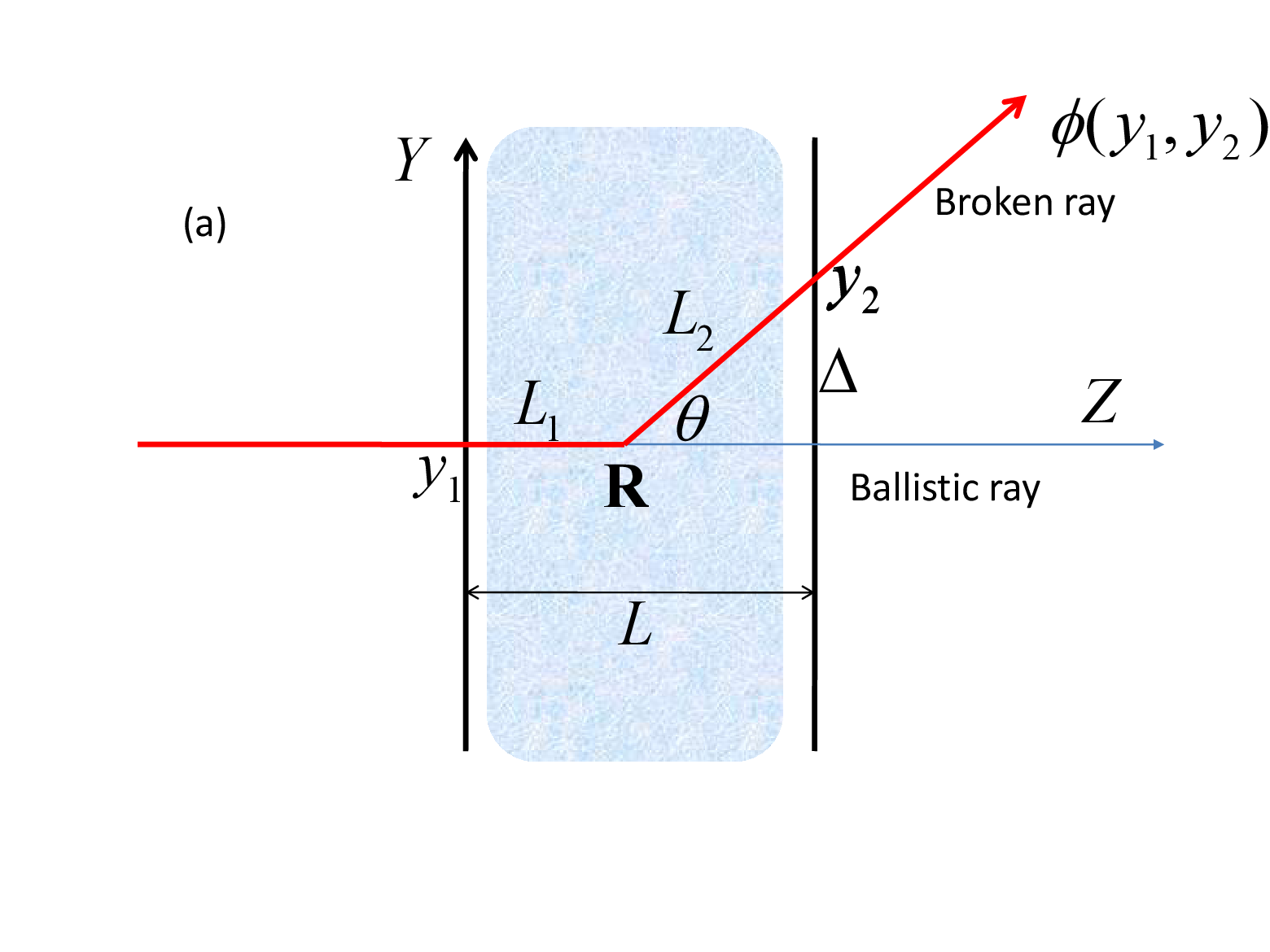,width=7cm,bbllx=50bp,bblly=100bp,bburx=720bp,bbury=540bp,clip=}
\psfig{file=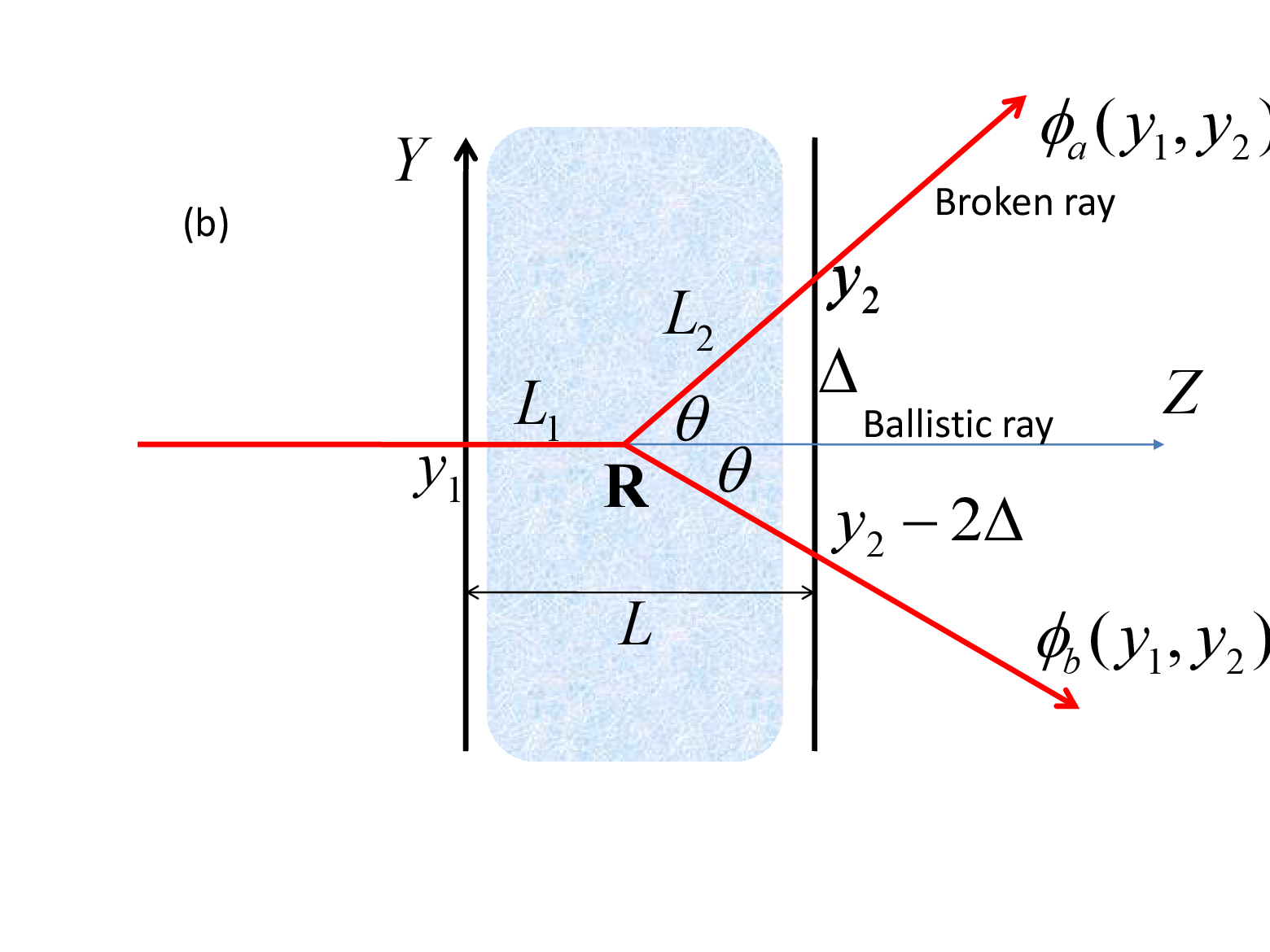,width=7cm,bbllx=50bp,bblly=100bp,bburx=720bp,bbury=540bp,clip=}
}
\caption{\label{fig1}
  (Color online) Geometry a broken ray for normal incidence.}
\end{figure}

The physics of broken ray formation has been discussed in
reference~\cite{florescu_09_1}. Here we focus on image reconstruction.
The geometry of a broken ray is illustrated in figure~\ref{fig1} for
the particular case of trans-illumination and normal incidence.

Imaging in SSOT is performed slice-by-slice. A slice is determined by
the direction vectors of sources and detectors which are assumed to be
sharply collimated and lie in the same plane. In the rectangular
reference frame of figure~\ref{fig1}, the slices correspond to
different planes $x={\rm const}$. Once a slice is selected, the
direction vectors are fixed, but the positions of the sources and
detectors can be scanned along the $Y$-axis, subject to the constraint
that the vertex ${\bf R}$ (the ray turning point) lies within the slab
$0<z<L$. The position of the source is denoted by $y_1$ and the
position of the detector by $y_2$, and every distinct source-detector
pair corresponds to a unique broken ray. The measured intensity of a
broken ray can be used to construct a {\em data function}
$\phi(y_1,y_2)$ as is described in reference~\cite{florescu_09_1}.
The attenuation and the scattering coefficients of the medium, $\mu_t$
and $\mu_s$, are related to the data function by the integral
transform

\begin{equation}
\label{Ray_int}
\int_{{\rm BR}(y_2,y_1)}\mu_t \big{(} {\bf r}(\ell) \big{)} d\ell
- \ln \frac{ \mu_s\big{(}{\bf R}(y_1,y_2) \big{)}}{\bar{\mu}_s}  = \phi(y_2,y_1) \ .
\end{equation}

\noindent
Here $\bar{\mu}_s$ is the background (average) scattering coefficient
of the medium, the integral is evaluated along the broken ray ${\rm
  BR}(y_2,y_1)$, ${\bf r}$ is a two-dimensional vector of position in
the $YZ$-plane, and ${\bf R}(y_1,y_2)$ is the ray turning point. The
problem of image reconstruction is to recover $\mu_t({\bf r})$ and
$\mu_s({\bf r})$ from a set of data points $\phi(y_2,y_1)$, whereas
the absorption coefficient $\mu_a({\bf r})$ can be determined as the
difference between $\mu_t({\bf r})$ and $\mu_s({\bf r})$.

If the scattering coefficient is spatially uniform and only absorptive
inhomogeneities are to be reconstructed, it is sufficient to use a
family of broken rays in which $y_2 > y_1$. We denote the transverse
source-detector separation by $\Delta$ ($\Delta = y_2 - y_1$). If the
angle between the illumination and detection directions is $\theta$,
then $\Delta$ varies in the interval $(0,\Delta_{\rm max})$, where
$\Delta_{\rm max} = L\tan\theta$.  The broken ray shown in
figure~\ref{fig1}(a) consists of two segments of length $L_1$ and
$L_2$, respectively, where

\begin{subequations}
\label{L12_def}
\begin{eqnarray}
\label{L1_def}
L_1(\Delta) = L - \Delta \cot\theta = L\left( 1 -
  \Delta/\Delta_{\rm max} \right) \ , \\ 
\label{L2_def}
L_2(\Delta) = \Delta \csc\theta  = \left( \Delta/\Delta_{\rm max} \right)
\sqrt{L^2 + \Delta_{\rm max}^2} \ .
\end{eqnarray}
\end{subequations}

If the absorption and the scattering coefficients can both fluctuate,
the measurements of the type schematically shown in
figure~\ref{fig1}(a) are insufficient to reconstruct the two
coefficients simultaneously. However, if the two broken rays with a
common vortex which are shown in figure~\ref{fig1}(b) are
registered, the problem becomes well-determined and the two
coefficients can be robustly recovered.

The measurement scheme described above is suitable for reconstructing
an image in the rectangular area defined by the inequalities $0<z<L$
and $y_{\rm min} < y < y_{\rm max}$, where $L$ is the depth of the
sample, and $y_{\rm min}$, $y_{\rm max}$ are determined by the window
in which the sources and detectors are scanned. In the case of optical
imaging, the measurement scheme described above can be experimentally
realized with the use of a CCD camera whose optical axis is tilted
with respect to the normal to the slab surface.

\section{Image reconstruction formulas}
\label{sec:img_rec}

\subsection{Spatially-uniform scattering}
\label{subsec:abs}

If the scattering coefficient is spatially-uniform, $\mu_s({\bf
  R})=\bar{\mu}_s$, and the logarithmic term in the left hand side of
equation~(\ref{Ray_int}) vanishes. The image reconstruction problem is
then reduced to recovering the function $\mu_t(y,z)$ from the integral
equation

\begin{equation}
\label{SSOT_1}
\int_{{\rm BR}(y_1,y_2)} \mu_t[y(\ell), z(\ell)] d\ell = \phi(y_1,y_2) \ .
\end{equation}

\noindent
It is convenient to introduce a change of variables, specifically,

\begin{equation}
\label{w_Delta}
y_1 = w \ , \ \ y_2 = w + \Delta \ ,
\end{equation}

\noindent
and define a new data function according to

\begin{equation}
\label{psi_phi}
\psi(w, \Delta) \equiv \phi(w,w+\Delta) \ . 
\end{equation}

\noindent
Thus, we parameterize the data by the position of the source, $w$, and
by the transverse source-detector separation, $\Delta$.  The shape of
a broken ray (within the slab) depends only on $\Delta$ but not on
$w$. Therefore, we can write

\begin{equation}
\label{eta_zeta_def}
y(\ell) = w + \eta(\Delta, \ell) \ , \ \ z(\ell) = \zeta(\Delta, \ell) \ ,
\end{equation}

\noindent
where the functions $\eta(\Delta, \ell)$, $\zeta(\Delta, \ell)$ are
independent of $w$. Taking a Fourier transform of equation~(\ref{SSOT_1})
with respect to $w$, we obtain a generalization of the Fourier-slice
theorem:

\begin{equation}
\label{psi_Fourier}
\int_0^{L_1(\Delta)+L_2(\Delta)} e^{ -i k \eta(\Delta,\ell)}
\tilde{\mu}_t \Big{(} k, \zeta(\Delta,\ell) \Big{)} d\ell =  \tilde{\psi}(k,\Delta) \ ,
\end{equation}

\noindent
where the Fourier transforms are defined as

\begin{equation}
\label{psi_mu_FT}
\tilde{\psi}(k,\Delta) = \int_{-\infty}^{\infty} \psi(w,\Delta) e^{i
  k w} d w \ , \ \
\tilde{\mu}_t(k,z) = \int_{-\infty}^{\infty} \mu_t(y,z) e^{i k y} dy \ .
\end{equation}

\noindent
Thus the two-dimensional integral equation (\ref{SSOT_1}) has been
reduced to the one-dimensional integral equation (\ref{psi_Fourier})
which is parameterized by the Fourier variable $k$.

The integral equation (\ref{psi_Fourier}) can be inverted
analytically. To this end, we must specify the functions
$\eta(\Delta,\ell)$ and $\zeta(\Delta,\ell)$. For the specific
geometry of the broken rays shown in figure~\ref{fig1}(a),

\begin{subequations}
\label{eta_zeta_norm}
\begin{eqnarray}
\label{eta_norm}
\hspace*{-2cm}\eta(\Delta, \ell) = \left \{
\begin{array}{ll}
0 \ , & \ \ell < L_1(\Delta) \\
\left[ \ell - L_1(\Delta) \right] \sin \theta \ , & \ L_1(\Delta)
< \ell < L_1(\Delta) + L_2(\Delta) 
\end{array} \right. , \\
\label{zeta_norm}
\hspace*{-2cm}\zeta(\Delta, \ell) = \left \{
\begin{array}{ll}
\ell \ , & \ \ell < L_1(\Delta) \\
L_1(\Delta) + \left[\ell - L_1(\Delta) \right] \cos \theta \ , &
\ L_1(\Delta) < \ell < L_1(\Delta) + L_2(\Delta) 
\end{array} \right. .
\end{eqnarray}
\end{subequations}

\noindent
Upon substitution of these expressions into (\ref{psi_Fourier}), we
obtain

\begin{equation}
\label{psi_Fourier_1}
\int_0^{L_1(\Delta)} \tilde{\mu}_t(k,\ell) d\ell 
+ \frac{e^{ik L_1(\Delta) \tan \theta}}{\cos \theta} \int_{L_1(\Delta)}^{L}
\tilde{\mu}_t(k,\ell) e^{-ik \ell \tan \theta} d\ell =
\tilde{\psi}(k,\Delta) \ .
\end{equation} 

\noindent
We can now use the degree of freedom associated with the variable
$\Delta$ to invert (\ref{psi_Fourier_1}) for any fixed value of $k$.
To simplify the derivations, we introduce several new notations:

\begin{subequations}
\begin{eqnarray}
& q    \equiv k\tan\theta \ , \\
& \lambda \equiv \cot(\theta/2) \ , \ \ c \equiv \cos\theta \ , \ \
\kappa = \frac{c}{1-c} =\cot(\theta/2)\cot\theta \ , \\
& f(z) \equiv \tilde{\mu}(q\cot\theta, z) \ , \ \
F(z) \equiv \tilde{\psi}\Big{(}q\cot\theta, (L-z)\tan\theta\Big{)} \ .
\end{eqnarray}
\end{subequations}

\noindent
The dependence of $f(z)$ and $F(z)$ on $q$ is implied. Then
(\ref{psi_Fourier_1}) takes the form

\begin{equation}
\label{psi_Fourier_2}
\int_{0}^{z} f(\ell) d\ell + \frac{1}{c} e^{iqz} \int_{z}^{L}
e^{-iq \ell} f(\ell) d\ell = F(z) \ , \ \ 0 \leq z \leq L \ .
\end{equation} 

\noindent
In this equation, $F(z)$ is known and $f(z)$ must be found.  To solve
(\ref{psi_Fourier_2}) for for a fixed value of $q$, we differentiate
once with respect to $z$ and obtain the following equation:

\begin{equation}
\label{app:20}
- \frac{1}{\kappa} f(z) + \frac{iq}{c} e^{iqz}
\int_z^L e^{-iq\ell} f(\ell) d\ell = F^\prime(z) \ ,
\end{equation}

\noindent
where prime denotes differentiation. We then use (\ref{psi_Fourier_2})
and (\ref{app:20}) to find the linear combination $G(z) = F^\prime(z)
- iq F(z)$ in terms of $f(z)$. The resultant equation is

\begin{equation}
\label{app:21}
-\frac{1}{\kappa}f(z) - iq \int_0^z f(\ell)d\ell = G(z) \ .
\end{equation}

\noindent
Differentiating one more time with respect to $z$, we obtain the
differential equation

\begin{equation}
\label{app:22}
f^\prime(z) + i \kappa q f(z) = - \kappa G^\prime(z) \ ,
\end{equation}

\noindent
which has the solution

\begin{equation}
\label{app:23}
f(z) = e^{-i\kappa q z} \left[ f(0) -\kappa \int_0^z e^{i\kappa q
    \ell} G^\prime(\ell)d\ell \right] \ . 
\end{equation}

\noindent
We then set $z=0$ in (\ref{app:21}) and find that $f(0)=- \kappa
G(0)$. Substituting this result into (\ref{app:23}) and integrating
once by parts, we arrive at the inverse solution to
equation~\ref{psi_Fourier_2}:

\begin{equation}
\label{inv_solution}
f(z) = - \kappa \left[ G(z) - i\kappa q e^{-i\kappa q z} \int_0^z
  e^{i\kappa q \ell} G(\ell)d\ell \right] \ .
\end{equation}

One important comment on the obtained solution is necessary. The
function $F(z)$ in (\ref{psi_Fourier_2}) is not arbitrary but such
that

\begin{equation}
\label{F_condition}
F(L) = e^{-i\kappa q L} \left[ cF(0) + i\kappa q \int_0^L
e^{i\kappa q \ell} F(\ell) d\ell \right] \ .
\end{equation}

\noindent
This can be verified directly. However, experimental measurements may
result in a function $F(z)$ that does not satisfy this condition. On
the other hand, the inverse solution (\ref{inv_solution}) is invariant
if we add to $F(z)$ a function of the form $a\exp(i q z)$, where $a$
is an arbitrary constant. It can be easily shown that any experimental
function $F(z)$ can be uniquely written in the form $F(z) = F_{\rm
  reg}(z) + a \exp(i q z)$, where $F_{\rm reg}(z)$ satisfies the
condition (\ref{F_condition}). Thus, inverse formula
(\ref{inv_solution}) involves regularization, or filtering of the
input data. 

Restoring the original notations, we find the inverse solution to
(\ref{psi_Fourier_1}):

\begin{equation}
\label{formula1}
\tilde{\mu}_t(k,z) = \lambda \left[ H(k,z) - ik\lambda e^{-ik\lambda z} 
\int_0^z e^{i\lambda k \ell} H(k,\ell) d\ell \right] \ ,  
\end{equation}

\noindent
where

\begin{equation}
\label{H_def}
H(k,z) \equiv \left. \left( \frac{\partial}{\partial \Delta} + ik
  \right) 
\tilde{\psi}(k,\Delta) \right\vert_{\Delta = (L-z)\tan\theta} \ .
\end{equation}

\noindent
The real space solution is obtained by applying the inverse Fourier
transform, namely,

\begin{equation}
\label{FT_inv}
\mu(y,z) = \int_{-\infty}^{\infty} \tilde{\mu}_t(k,z) e^{-i k y} \frac{dk}{2\pi} \ ,
\end{equation}

\noindent
which yields

\begin{eqnarray}
\label{formula1_realspace}
\mu_t(y, z) &=&  \lambda \left\{ \left[ \frac{\partial}{\partial\Delta} -
    (1+\kappa)\frac{\partial}{\partial y} \right] \psi(y, \Delta) + \kappa
  \frac{\partial}{\partial y} \psi(y + \lambda z, \Delta_{\rm
      max} ) \right.
\nonumber \\
&-& \left.\left. \kappa(1+\kappa) \frac{\partial^2}{\partial y^2}
\int_{\Delta}^{\Delta_{\rm max}} 
\psi\Big{(}y + \kappa(\ell - \Delta) , \ell \Big{)} d\ell
\right\} \right\vert_{\Delta = (L-z)\tan\theta} \ .
\end{eqnarray}

\noindent
The above equation is the generalization of the conventional filtered
backprojection formula to the case of the broken ray transform
(\ref{SSOT_1}).

\subsection{Spatially-nonuniform scattering}
\label{subsec:abs+scat}

Simultaneous reconstruction of scattering and absorption can be
realized without making use of the entire parameter space which is
available in SSOT. It is sufficient, for example, to use normal
incidence and two different angles of detection. A particular case of
this measurement scheme is illustrated in figure~\ref{fig1}(b),
where the source is scanned along the $Y$-axis and the intensity of
two distinct broken rays is measured for every position of the source.
The two broken rays have a common vortex at the point ${\bf R}$ and
are denoted as ${\rm BR}_a$ and ${\rm BR}_b$. The data functions
obtained by measuring the intensities of respective rays are denoted
by $\phi_a(y_1,y_2)$ and $\phi_b(y_1,y_2)$.

At the first step, we eliminate the logarithmic term in
equation~\ref{Ray_int} by taking a differential measurement.  The
differential data function is defined as

\begin{equation}
\label{phi_d_def}
\phi_d(y_1,y_2) = \phi_a(y_1,y_2) - \phi_b(y_1,y_2) \ .
\end{equation}

\noindent
It is easy to see that the $\phi_d(y_1,y_2)$ can be used to
reconstruct $\mu_t(y,z)$ even when the scattering coefficient of the
medium is spatially-nonuniform. Indeed, because the a- and b-type
broken rays have the same vortex, we have

\begin{equation}
\label{SSOT_2D}
\int_{{\rm BR}_a(y_1,y_2)} \mu_t\big{(}y(\ell), z(\ell)\big{)} d\ell -
\int_{{\rm BR}_b(y_1,y_2)} \mu_t\big{(}y(\ell), z(\ell)\big{)} d\ell = 
\phi_d(y_1,y_2) \ , 
\end{equation}

As above, the inversion formula for (\ref{SSOT_2D}) can be derived by
employing the Fourier transform. We make the change of variables
(\ref{w_Delta}) and define the new data function as $\psi_d(w,\Delta)
\equiv \phi_d(w,w+\Delta)$. The Fourier slice theorem then takes the
form

\begin{equation}
\label{psi_Fourier_inhom}
-2i\int_{L_1(\Delta)}^{L_1(\Delta)+L_2(\Delta)} \sin\Big{(} k \eta(\Delta, \ell) \Big{)}
\tilde{\mu}_t \Big{(} k, \zeta(\Delta, \ell) \Big{)} d\ell =
\tilde{\psi}_d(k,\Delta) \ ,
\end{equation}

\noindent
where $\eta (\Delta, \ell)$ and $\zeta(\Delta, \ell)$ are given by
(\ref{eta_zeta_norm}). Using these expressions, and making an
appropriate change of the integration variable,
(\ref{psi_Fourier_inhom}) can be re-written as

\begin{equation}
\label{psi_k}
-\frac{2i}{\cos \theta} \int_{L_1(\Delta)}^{L} \sin \Big{(} k ( \ell-L_1(\Delta) )\tan
\theta\Big{)} \tilde{\mu}_t(k, \ell) d\ell = \tilde{\psi}_d(k,\Delta) \
. 
\end{equation}

\noindent
To solve for $\tilde{\mu}_t(k, z)$, we use the change of variables
$z=L_1(\Delta)$, (equivalently, $\Delta=(L-z)\tan\theta$) and
differentiate twice with respect to $z$. This yields

\begin{equation}
\label{mu_k2}
\tilde{\mu}_t(k, z)=\left.\frac{\sin \theta}{2}
\left(-\frac{1}{ik}\frac{\partial^2 }{\partial \Delta^2} + ik\right) 
\tilde{\psi}_d(k,\Delta) \right \vert_{\Delta = (L-z)\tan\theta} \ .
\end{equation}

\noindent
The real-space solution $\mu_t(y,z)$ is obtained by the inverse
Fourier transform. The final result is

\begin{equation}
\mu_t(y,z) = \left. \frac{\sin\theta}{4}
\left\{ \frac{\partial^2}{\partial \Delta^2} \int_{-\infty}^{\infty}
{\rm sgn}(y-w) \psi_d(w,\Delta) dw  - 2
\frac{\partial}{\partial y} \psi_d(y,\Delta) \right] \right
\vert_{\Delta = (L-z)\tan\theta} \ .
\label{formula2}
\end{equation}

\noindent
In the above equation, ${\rm sgn}(x)$ denotes the sign of $x$ and it
is assumed that the data function $\psi_d(w,\Delta)$ vanishes for
$\vert w \vert > w_{\rm max} >0$, so that the integral in the
right-hand side of (\ref{formula2}) converges.

With $\mu_t(y,z)$ thus determined, $\mu_s(y,z)$ can be obtained
directly from (\ref{Ray_int}), where one of the two rays (either the
a-type or the b-type) is used. Finally, the absorption coefficient in
obtained as $\mu_a(y,z)= \mu_t(y,z)-\mu_s(y,z)$.

\section{Numerical Simulations}
\label{sec:num}

In this section, we illustrate the inversion formulas derived above
with numerical examples. In the case of a constant scattering
coefficient, we will use the Fourier-space formulas (\ref{formula1}),
(\ref{H_def}) and then the inverse Fourier transform (\ref{FT_inv}) to
reconstruct the total attenuation coefficient $\mu_t(y,z)$. In the
case of spatially-varying $\mu_s$, the real-space formula
(\ref{formula2}) will be used to reconstruct $\mu_t(y,z)$, then the
result will be substituted into (\ref{SSOT_2D}) to reconstruct
$\mu_s(y,z)$; the absorption coefficient will be obtained as $\mu_a =
\mu_t - \mu_s$. In all cases, the reconstructions are carried out in a
rectangular area $0<z<L$, $0<y<3L$ and the detection angle is
$\theta=\pi/4$ so that $\Delta_{\rm max} = L$.

In what follows, the optical coefficients are decomposed as

\begin{equation}
\label{mu_decomp}
\mu_t(y,z) = \bar{\mu}_t + \delta\mu_t(y,z) \ , \ \ \mu_s(y,z) =
\bar{\mu}_s + \delta\mu_s(y,z) \ , \ \ \mu_a(y,z) = \bar{\mu}_a +
\delta\mu_a(y,z) \ .
\end{equation}

\noindent
Here the quantities with a bar are constant background values of the
respective coefficients and the $\delta$-terms represent the
inhomogeneities. In performing image reconstruction, it will be
assumed that the support of the functions $\delta\mu_a(y,z)$ and
$\delta\mu_s(y,z)$ is completely contained in the rectangular area
where the image is reconstructed. This assumption is, strictly
speaking, violated if Gaussian targets are used, as will be done
below. However, the Gaussian inhomogeneities are exponentially small
outside of the imaging area, and the error incurred due to the
inaccurate assumption is also exponentially small.

\subsection{Spatially-uniform scattering}

We consider first the case of a medium with $\delta\mu_s = 0$ and a
spatially-inhomogeneous attenuation coefficient (due to
spatially-varying absorption) in the shape of a square with sharp
boundaries, viz,

\begin{equation}
\label{square}
\delta\mu_t(y,z) = \left \{
\begin{array}{ll}
\bar{\mu}_t \ , & \ \vert y-y_0 \vert \le a/2 \ \mbox{and} \ \vert z - z_0\vert \le a/2 \\
0 \ , & \mbox {otherwise}
\end{array}\right. \ .
\end{equation}

\noindent
Thus, the attenuation coefficient inside the square is twice the
background value of $\bar{\mu}_t$. The square is centered so that
$y_0=L$, $z_0=L/2$ and its side length is $a=L/2$. The Fourier-space
data function $\tilde{\psi}(k,\Delta)$ was obtained analytically from
(\ref{formula1}) by direct integration. Then the variables $\Delta$
and $k$ were sampled and $\mu_t(y,z)$ was reconstructed using discrete
samples of $\tilde{\psi}(k,\Delta)$. More specifically, the
samples have been used to evaluate numerically the equations
(\ref{FT_inv}) and (\ref{formula1}). The variable $\Delta$ was sampled
as $\Delta_n = hn$, where $n=0,1,\ldots N$ and $h=\Delta_{\rm max}/N$.
Note that for $\theta=\pi/4$, $\Delta_{\rm max}=L$. The Fourier
variable $k$ was sampled in the interval $[-\pi/h,\pi/h]$ according to
$k_m = (\pi/h)(m/N - 1)$, where $m=0,1,\ldots,2N$. The derivative in
(\ref{H_def}) was computed as the central difference

\begin{equation}
\label{finite_diff_def}
\left. \frac{\partial \tilde{\psi}(k,\Delta)}{\partial \Delta}
\right\vert_{\Delta = \Delta_n} = \frac{\tilde{\psi}(k,\Delta_{n+1}) -
  \tilde{\psi}(k,\Delta_{n-1})}{2h} \ , \ \ n=0,1,\ldots, N \ .
\end{equation}

\begin{figure}
\centerline{
\psfig{file=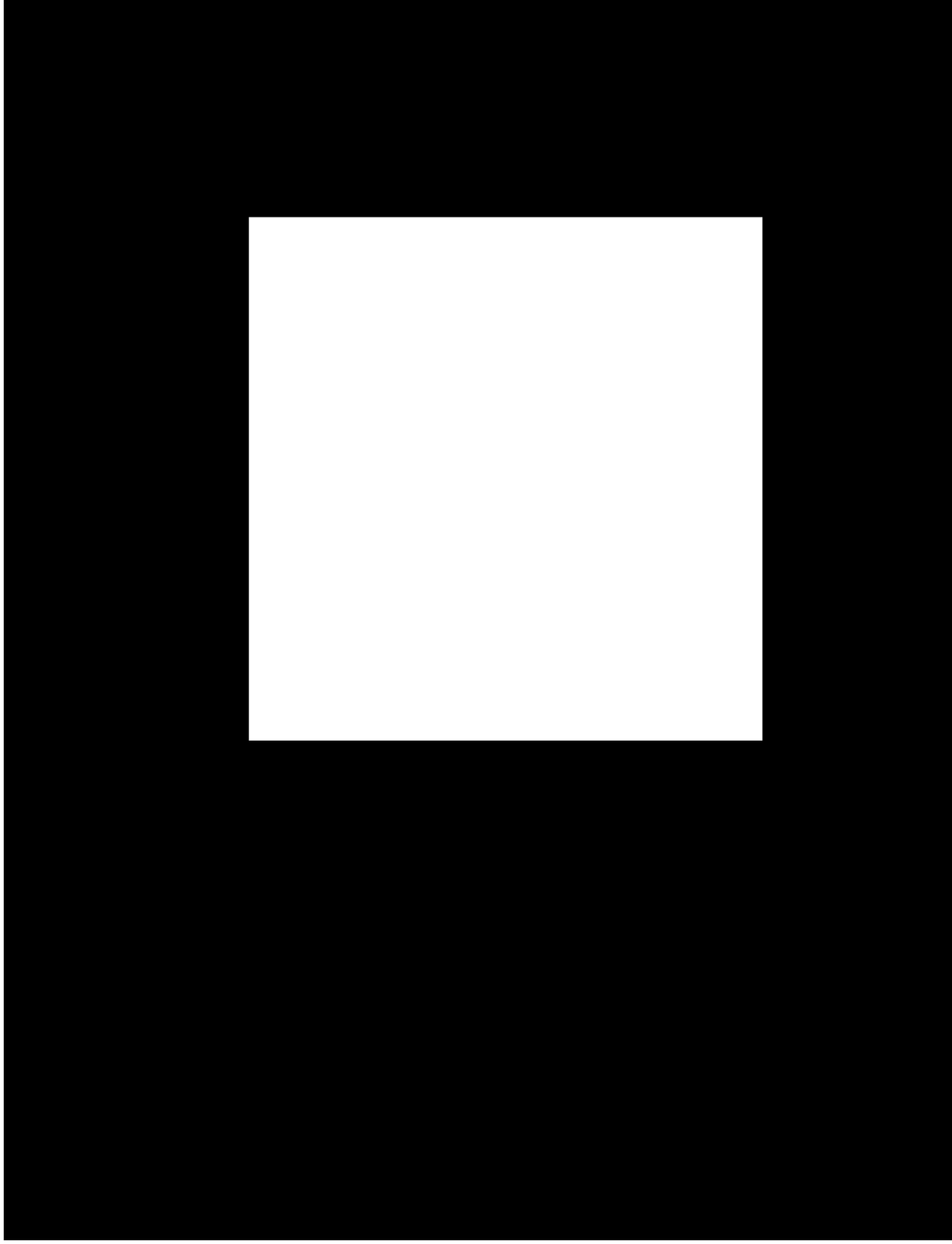,height=7cm}
\psfig{file=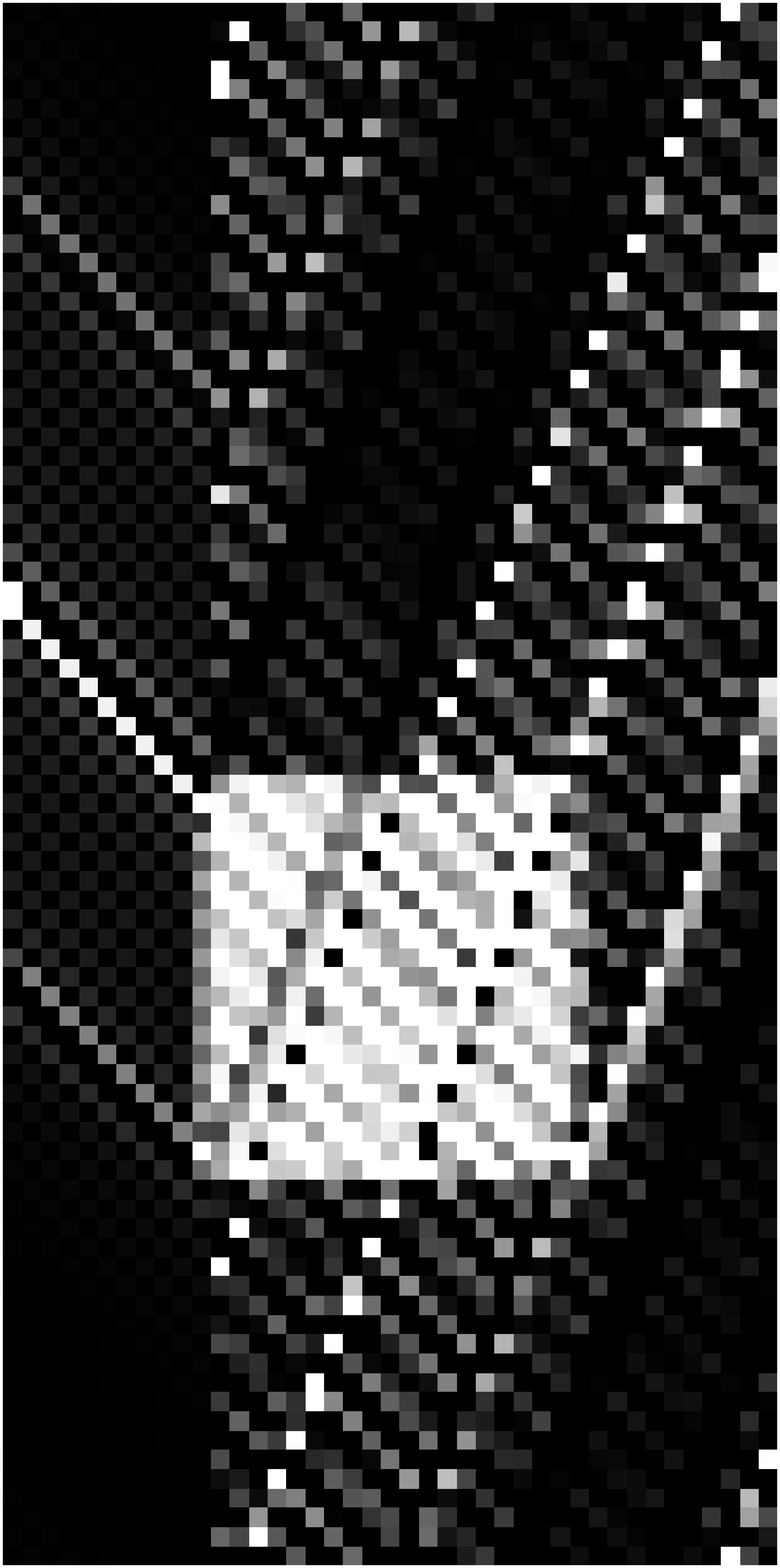,height=7cm}
\psfig{file=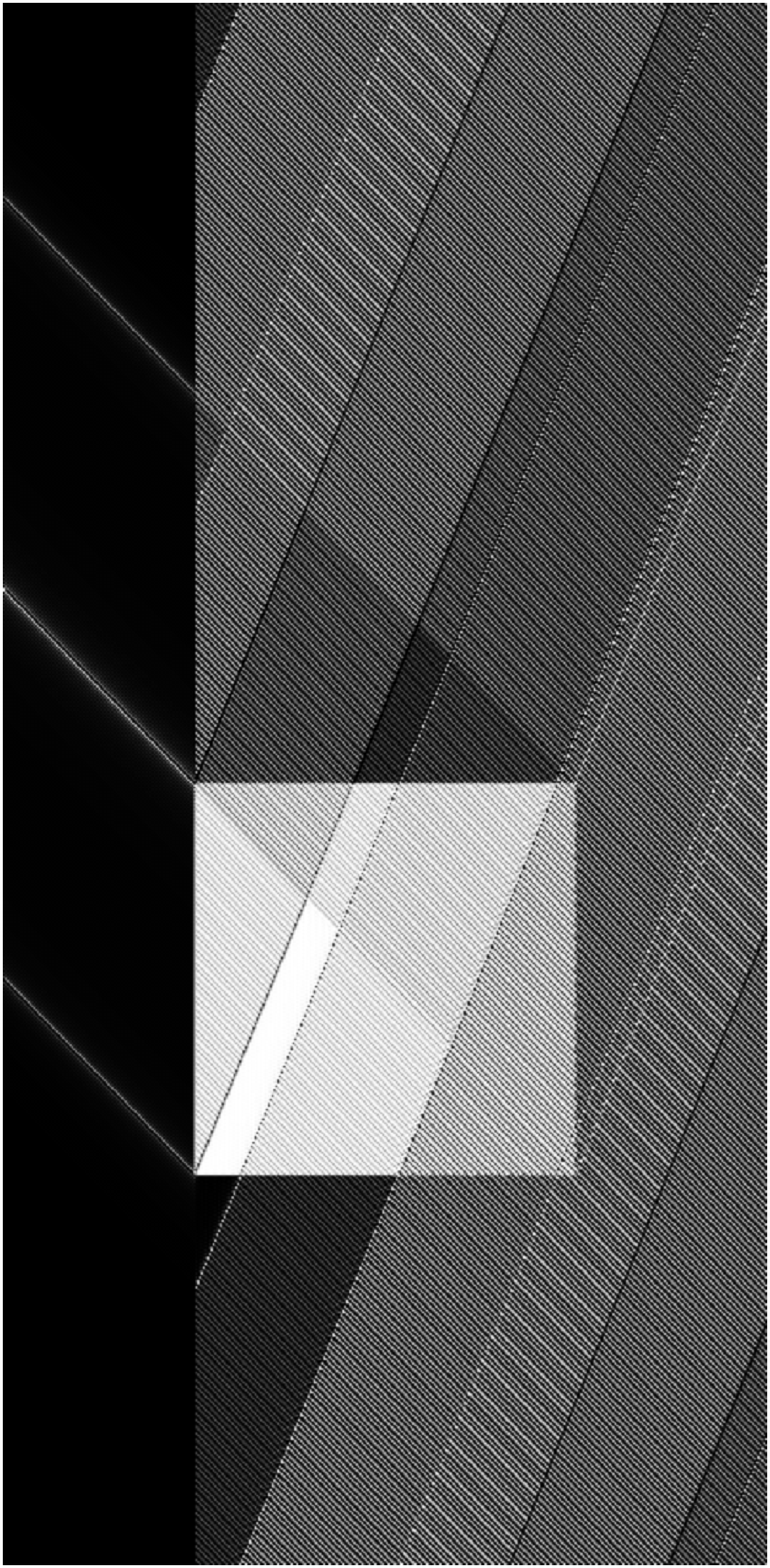,height=7cm}
}
\centerline{Model \hspace*{2cm} $N=40$ \hspace*{2cm} $N=400$}
\caption{\label{fig2}
  Reconstructed $\delta\mu_t(y,z)$ for the square target
  (\ref{square}) using different number of samples $N$, as labeled.}
\end{figure}

\noindent
In applying the above formula, the boundary condition
$\tilde{\psi}(k,\Delta_{-1}) = \tilde{\psi}(k,\Delta_{N+1})$ has been
used. Note that the substitution $\Delta \rightarrow (L-z)\tan\theta$
contained in the formula (\ref{H_def}) did not require any re-sampling
because, in the geometry used, $\tan\theta=1$. More generally,
however, re-sampling and interpolation is required to apply
(\ref{H_def}) to a uniformly sampled function
$\tilde{\psi}(k,\Delta)$. The integral over $\ell$ in (\ref{formula1})
was evaluated numerically using the trapezoidal rule.  Finally, the
image was reconstructed on a rectangular grid with the same step as
was used to sample the variable $\Delta$, that is, $h$.
Reconstructions of the total attenuation coefficient $\mu_t(y,z)$
obtained as described above is shown in figure~\ref{fig2} for two
different values of the parameter $N$. It can be seen that the
reconstruction contains artifacts. When the number of samples, $N$, is
increased by the factor of $10$, the support of the artifacts is
reduced by the same factor, but the amplitude stays unchanged. We have
verified that the $L^2$ norm of the discrepancy $\xi(y,z) = \mu_t^{\rm
  (true)}(y,z) - \mu_t^{\rm (reconstructed)}(y,z)$ tends to zero when
$N\rightarrow \infty$, yet the maximum amplitude of the relative
error, $\max [\xi^{\rm (true)}(y,z)/\mu_t(y,z)]$, remains of the order
of unity.

The relatively low image quality seen in figure~\ref{fig2} is
caused by the sharp discontinuity of the target. This statement is
confirmed by considering a smooth target in which the attenuation
coefficient is given by a Gaussian, viz,

\begin{equation}
\label{Gaussian}
\delta\mu_t(y,z)=\bar{\mu}_t \exp\left(-\frac{(y-y_0)^2 +
    (z-z_0)^2}{\sigma^2} \right) \ .
\end{equation}

\begin{figure}
\centerline{\psfig{file=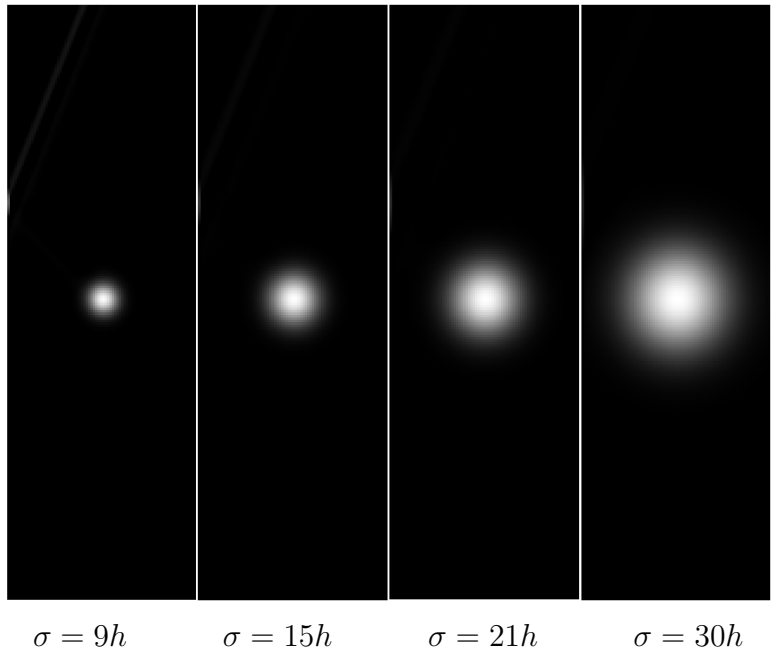,height=8cm}}
\centerline{$\sigma=9h$ \hspace*{1.0cm} $\sigma=15h$ \hspace*{1.0cm}
  $\sigma=21h$ \hspace*{1.0cm} $\sigma=30h$}
\caption{\label{fig3}
  Reconstructed $\delta\mu_t(y,z)$ for the Gaussian target
  (\ref{Gaussian}) using $N=120$ and different values of $\sigma$, as
  labeled.}
\end{figure}

\begin{figure}
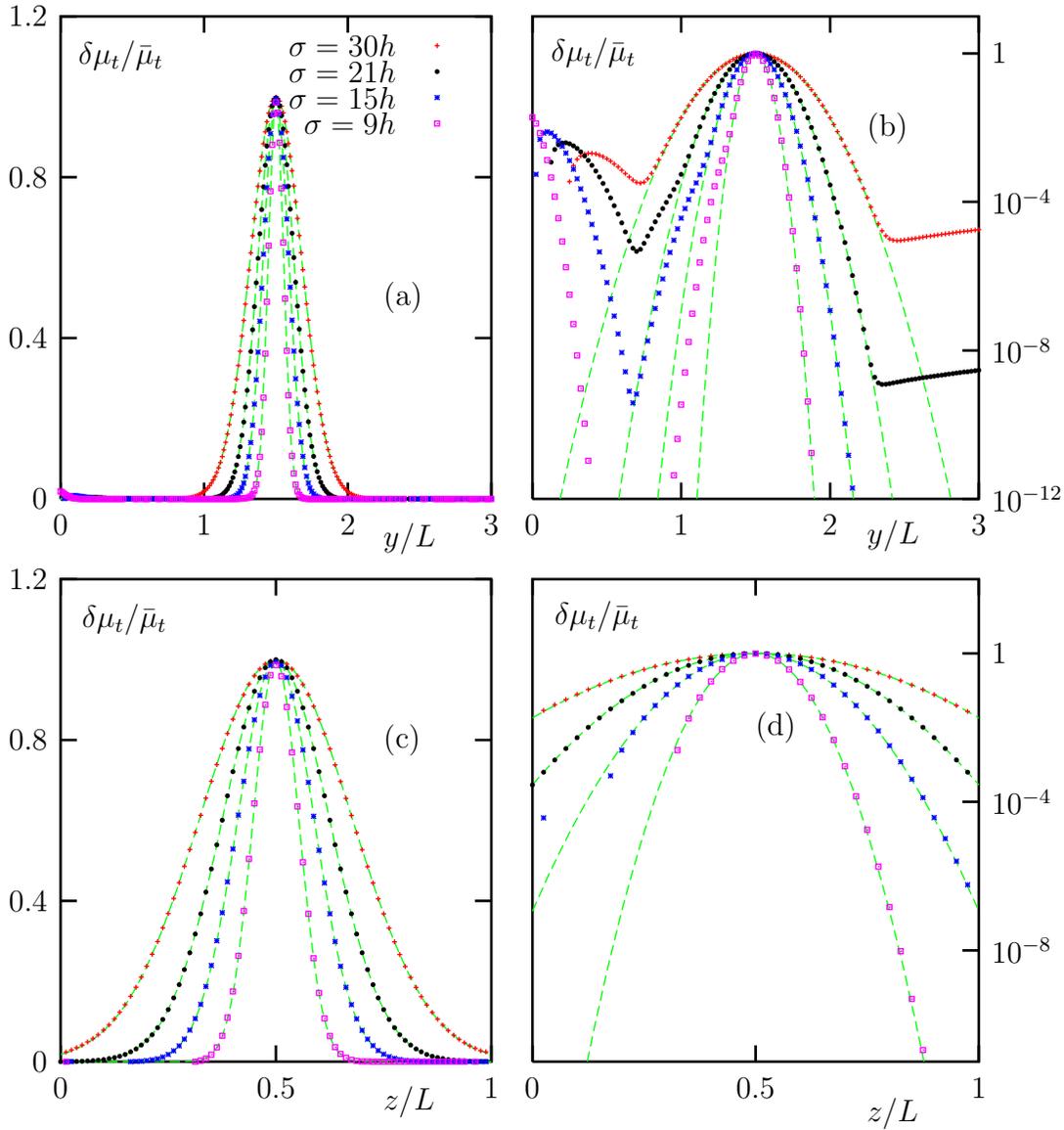

\input{Fig4a.tex}\hspace*{-5mm}\input{Fig4b.tex}
\input{Fig4c.tex}\hspace*{-5mm}\input{Fig4d.tex}
\caption{\label{fig4}
  (Color online) The data of figure~\ref{fig3} are shown here as cross
  sections along the straight lines drawn through the center of the
  inhomogeneity in the directions parallel to the $Y$-axis (a,b) and
  to the $Z$-axis (c,d). The panels (a,c) use the linear scale and the
  panels (b,d) use the logarithmic scale for the vertical axis. Dots
  represent the reconstructed values and the model function is
  represented by the dashed lines. Every second reconstructed data
  point is shown in semi-logarithmic plots. Some of the data points
  are not shown in the semi-logarithmic plots because the respective
  values are negative or too small to be displayed.}
\end{figure}

\noindent
The data function for this target, $\tilde{\psi}(k,\Delta)$, can be
computed analytically (the result contains the error function). We
have used the same sampling procedure as above and a discretized image
of the target (\ref{Gaussian}) has been reconstructed using equations
(\ref{H_def}),(\ref{formula1}) and (\ref{FT_inv}). The results are
shown in figures~\ref{fig3} and \ref{fig4}. It can be seen that an
accurate quantitative reconstruction is obtained for the smooth target
of the type (\ref{Gaussian}). When the linear scale is used to
represent the data, no artifacts are visible in the reconstructions
and the reconstructed data coincide quantitatively with the model.
However, when the logarithmic scale is used on the vertical axis, as
is done in panels (b,d) of figure \ref{fig4}, the artifacts become
clearly visible. The amplitude of the artifacts in the lateral cross
section of the image (figure \ref{fig4}(b)) is, in fact, less than 2\%
of the plot maximum.

\subsection{Spatially-nonuniform scattering}

We next consider the case when both the scattering and the absorption
coefficients of the medium are varying. We have modeled the
inhomogeneities as Gaussians

\begin{equation}
\label{Gaussian_sa}
\delta\mu_{s,a}(y,z)=\bar{\mu}_{s,a} \exp\left(-\frac{(y-y_{s,a})^2 +
    (z-z_{s,a})^2}{\sigma^2} \right) \ .
\end{equation}

\noindent
where $(y_s, z_s)$ and $(y_a, z_a)$ are the centers of the scattering
and the absorbing inhomogeneities. Note that the amplitude of each
coefficient in the center of an inhomogeneity is twice the background
value.  Both kinds of inhomogeneity were assumed to be present in the medium
simultaneously but not overlap. Thus, the scattering inhomogeneity was
centered at the point $(y_s=3.125L, z_s=0.5L)$ and the absorbing
inhomogeneity was centers at $(y_a=0.875L, z_a=0.5L)$. We have
considered different values of the parameter $\sigma$. However, in
every reconstruction, $\sigma$ was the same for the absorbing and the
scattering inhomogeneity. 

In this subsection, we have used the real-space image reconstruction
formula (\ref{formula2}). The data function was obtained by analytical
integration and the image reconstruction was performed by sampling the
variables $w$ and $\Delta$ in the data function $\psi_d(w,\Delta)$ on
a rectangular grid with the step $h$, as is described in more detail
in the previous subsection. In the reconstructions of this subsection,
$N=120$ and $h=L/N$. The derivatives in (\ref{formula2}) were computed
by central differences and the integral by the trapezoidal rule.

\begin{figure}
\centerline{\psfig{file=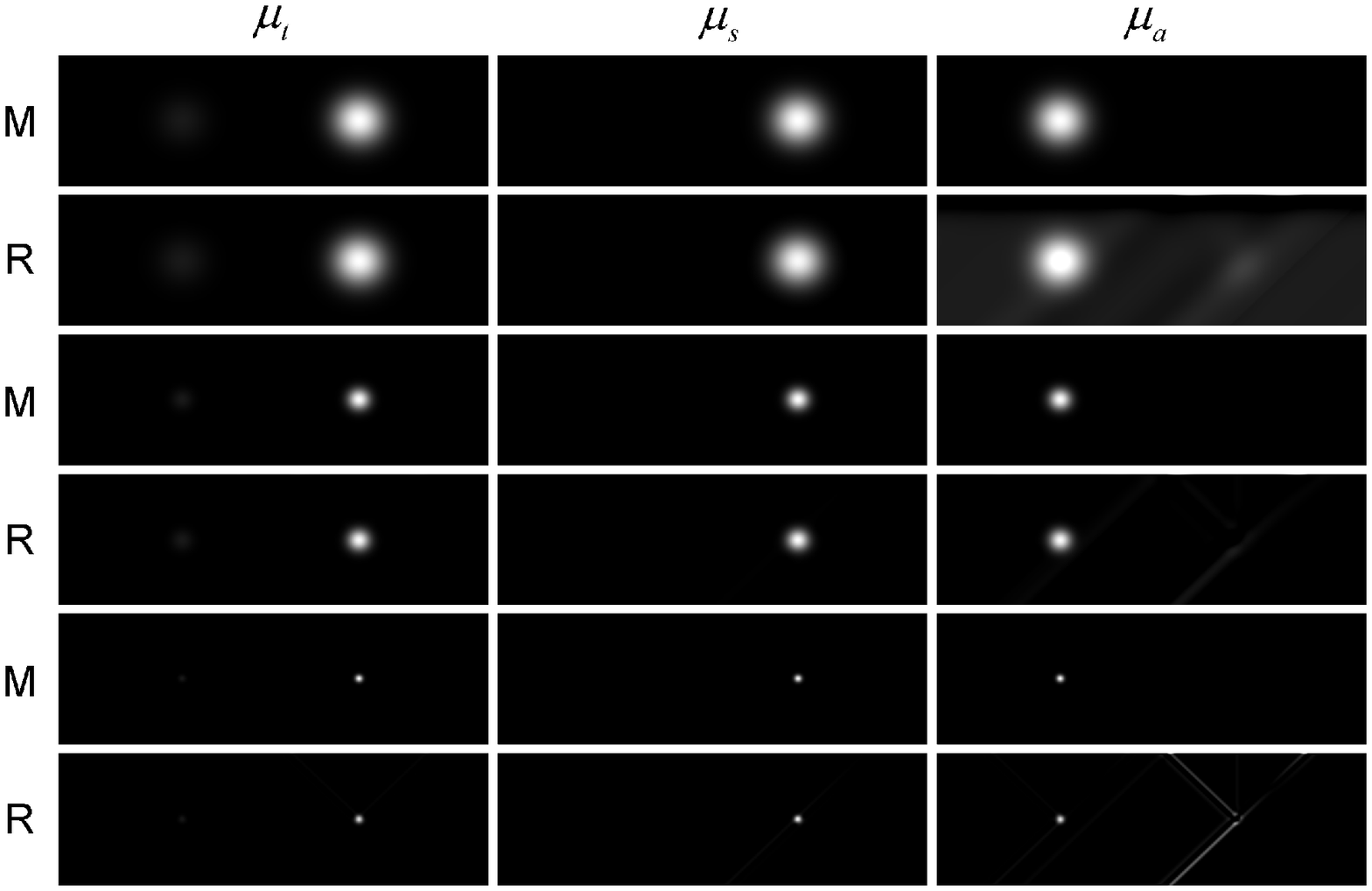,height=10cm}}
\caption{\label{fig5}
  Simultaneous reconstruction of the absorbing and scattering
  coefficients for the case $\bar{\mu}_s L = 2.4$, $\bar{\mu}_a
  L=0.24L$. The three columns represent the attenuation, scattering
  and absorption coefficients, as labeled.  The letter ``M'' indicates
  ``model'' and the letter ``R'' indicates reconstruction. The first
  two rows correspond to $\sigma=21h$, the next two rows correspond to
  $\sigma=9h$ and the last two rows correspond to $\sigma=3h$. Here
  $h=L/N$ and $N=120$. Every plot is normalized to its own maximum.}
\end{figure}

\begin{figure}
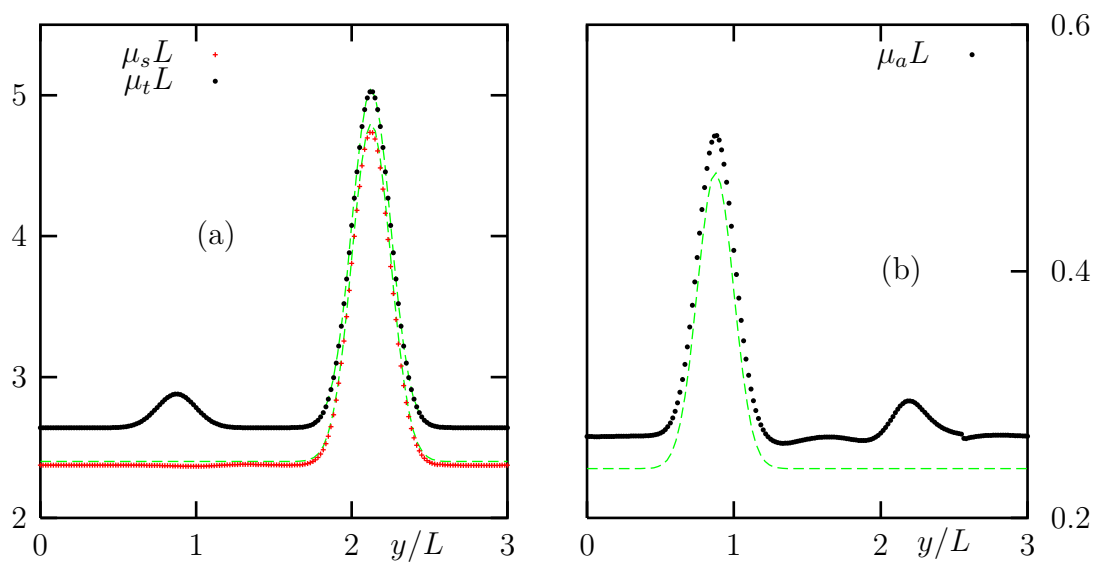

\input{Fig6a.tex}\input{Fig6b.tex}
\caption{\label{fig6}
  (Color online) The data of figure~\ref{fig5} for the case
  $\sigma=21h$ ($h=L/N$ and $N=120$) are shown here as cross sections
  along the straight line $z=z_a=z_s$ which intersects the centers of
  the absorbing and the scattering inhomogeneities. Total attenuation
  and scattering are shown in panel (a) and absorption is shown in
  panel (b). Centered symbols correspond to the reconstructed values
  and the dashed lines correspond to the model. Every second
  reconstructed data point is shown.}
\end{figure}

\begin{figure}
\centerline{\psfig{file=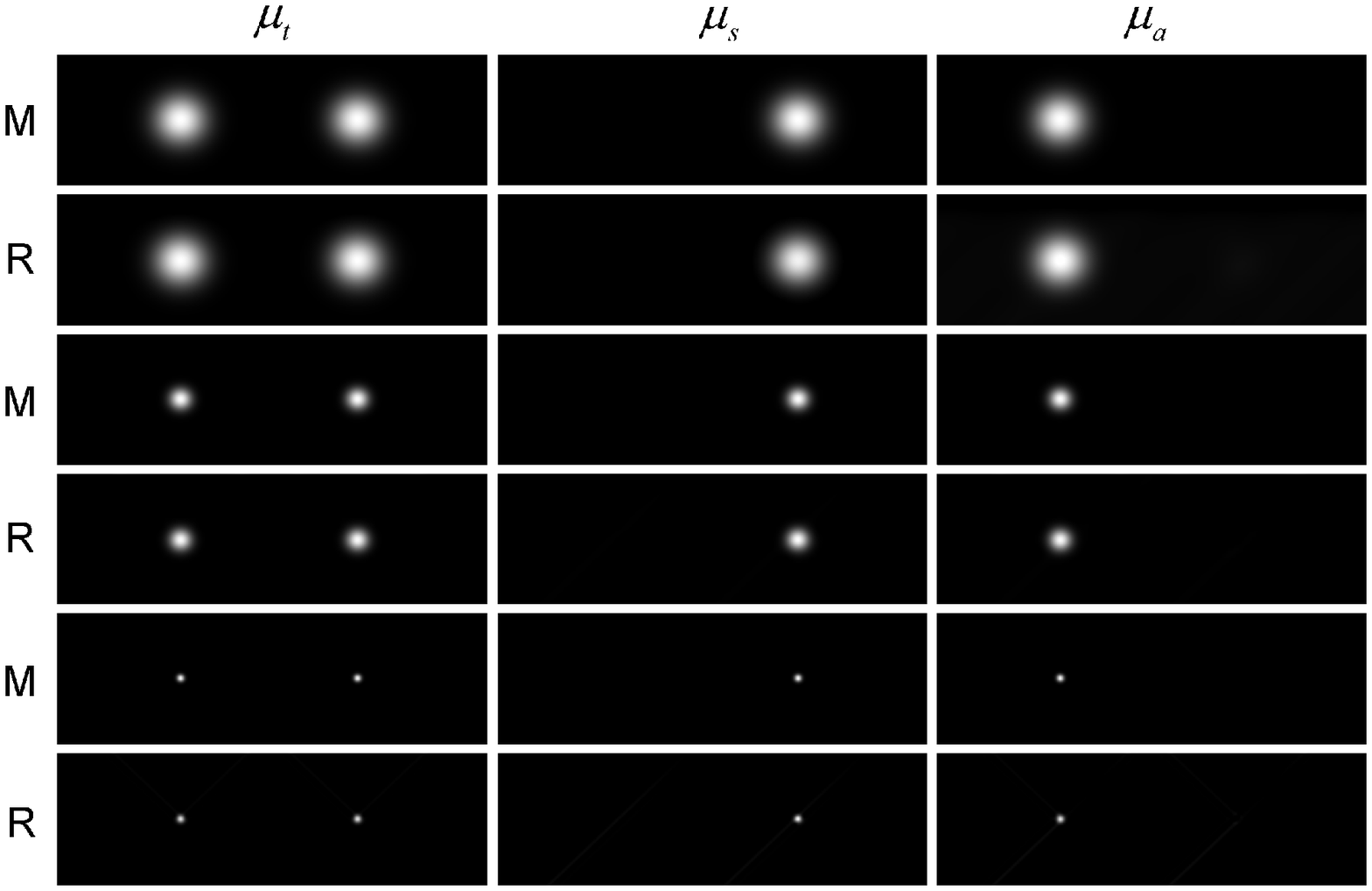,height=10cm}}
\caption{\label{fig7}
  Same as for Fig.~\ref{fig5} but for $\bar{\mu}_s L =\bar{\mu}_a L = 2.4$.}
\end{figure}

\begin{figure}
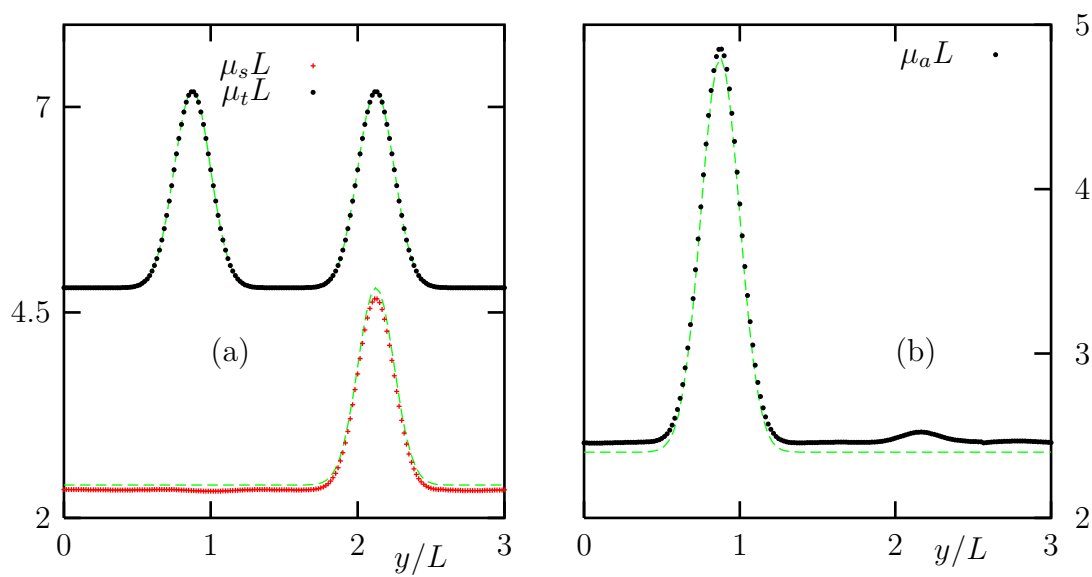

\input{Fig8a.tex}\input{Fig8b.tex}
\caption{\label{fig8}
  (color online) Same as for Fig.~\ref{fig6} but for $\bar{\mu}_s L
  =\bar{\mu}_a L = 2.4$.}
\end{figure}

We consider below two cases. The first case corresponds to
$\bar{\mu}_s L = 2.4$ and $\bar{\mu}_a L=0.24$, so that the
background scattering coefficient is ten times smaller than the
background scattering coefficient. The challenge here is to
reconstruct the absorbing inhomogeneities in the presence of much
stronger scattering inhomogeneities. In the second case, $\bar{\mu}_s
L = \bar{\mu}_a L = 2.4$, so that the strength of the absorbing and
the scattering inhomogeneities is the same. Image reconstruction for
the first case is illustrated in figures~\ref{fig5} and \ref{fig6}.
Here we plot total optical coefficients (including the background)
rather than the fluctuating parts $\delta\mu_t$, etc., as was done in
figures~\ref{fig2},\ref{fig3},\ref{fig4}. It can be seen that a good
image quality and a quantitative agreement with the model are obtained
for the scattering and the attenuation coefficients. The
reconstruction of the absorption coefficient is not as good. This is
because the relatively small quantity $\mu_a$ was obtained by finding
the numerical difference between the two much larger quantities
$\mu_t$ and $\mu_s$. In the case when the magnitudes of the scattering
and the absorbing inhomogeneities are the same, a quantitatively
accurate reconstruction of all three coefficients is obtained, as is
illustrated in figures~\ref{fig7} and \ref{fig8}.

\section{Conclusion}
\label{sec:sum}

We have derived and tested numerically image reconstruction formulas
for the broken ray transform (also referred to as the V-line
transform). The obtained formulas are generalizations of the filtered
backprojection formula of the conventional Radon transform. We have
shown that the broken ray transform is in certain aspects more useful
as it allows one to reconstruct the scattering and the absorption
coefficients of the medium simultaneously.

Inversion of the broken ray transform is mildly ill-posed. The
ill-posedness strongly affects image reconstruction when the target
has sharp boundaries. We conjecture that the ill-posedness can be
regularized by using additional rays.

The broken ray transform considered in this paper may become useful on
its own merits in the situations when scattering in the medium is not
negligible. However, it can also prove useful in conjunction with the
conventional Radon transform in which the intensity of ballistic rays
is measured. Since the ballistic ray is a special case of the broken
ray, it seems plausible that a reconstruction formula can be obtained
which would utilize the measurements of ballistic and broken rays
simultaneously. This will be the subject of our future work.

\section*{Acknowledgment}
This work was supported by the NSF under Grants No. DMS-0554100 and
No.  EEC-0615857, and by the NIH under Grant No.  R01EB004832.  The
authors express their deep gratitude to Plamen Stefanov, Guillaume
Bal, Peter Kuchment and Lihong Wang for stimulating and illuminating
discussions.

\section*{References}

\bibliographystyle{iopart-num} 
\bibliography{abbrev,local,master}

\end{document}